\documentclass[notitlepage,11pt,oneside]{article}
\usepackage[footnotesize,bf,hang]{caption}
\usepackage{verbatim}
\usepackage{epsfig}
\usepackage{amssymb}
\usepackage[dvips]{color}
\usepackage{amsmath}
\usepackage{amsfonts}
\usepackage{enumerate}
\usepackage{calc}
\usepackage{graphics}

\usepackage{subfigure}
\usepackage[french,english]{babel}

\begin{document}
\vspace{-1cm}
\enlargethispage*{3\baselineskip}
\title{\vspace{-1.5cm}The controversial piston 
in the thermodynamic limit\vspace{-0.4cm}} 

\vspace{-0.3cm}\author{Christian Gruber  and S\'everine Pache\\{\it \footnotesize
Institut de Physique Th\'{e}orique, Ecole Polytechnique
F\'ed\'erale de Lausanne, CH-1015 Lausanne, Switzerland} }

\vspace{-0.2cm}\date{\vspace{-0.3cm}\small\today}\vspace{-1.1cm}
\maketitle
\vspace{-1.3cm}
\section*{\centering\small Abstract}
\vspace{-0.2cm}{\small
\indent\indent We consider the evolution of a system composed of $N$ non-interacting point particles of mass $m$ in a container divided in two regions by a movable adiabatic wall (adiabatic piston). In this talk we  discuss the thermodynamic limit where the area $A$ of the container , the number $N$ of  particles, and  the mass $M$  of the piston go to infinity keeping $\frac{A}M $ and $\frac{N}M $ fixed. We  show that in this limit the motion of the piston is deterministic. Introducing simplifying assumptions we discuss the approach to equilibrium and we illustrate the results with numerical simulations. The comparison with the case of a system with finite $(A, N, M)$ will be presented.
\
\\ \vspace{-0.2cm} \\
\noindent\noindent PACS: 05.20 Dd;  05.40.+j; 05.70.Ln; 02.50.+s

\noindent\noindent{\it Keywords:} Liouville equation, adiabatic, piston, mechanical equilibrium, thermal equilibrium, damping.}\vspace{-0.2cm}
\section{Introduction}
\indent\indent The "adiabatic piston" is a well known problem in thermodynamics. Two gases enclosed in a cylinder are separated by an adiabatic movable piston. Initially the piston is at rest at the position $X_0$
and the two gases on the left $(-)$ and on the right $(+)$ of the piston are in thermal equilibrium with pressure $p^{\pm}$ and temperature  $T^{\pm}$. At time $t=0 $  the piston is let free to move and the problem is to find the final equilibrium state, $i.e.$ the position $X_f$ of the piston and the state $(p_f^{\pm}, T_f^{\pm})$ of the two gases $(fig. 1.a)$.
\\ 
 \begin{center}
\includegraphics[width=13.5 cm,height=3.5 cm]{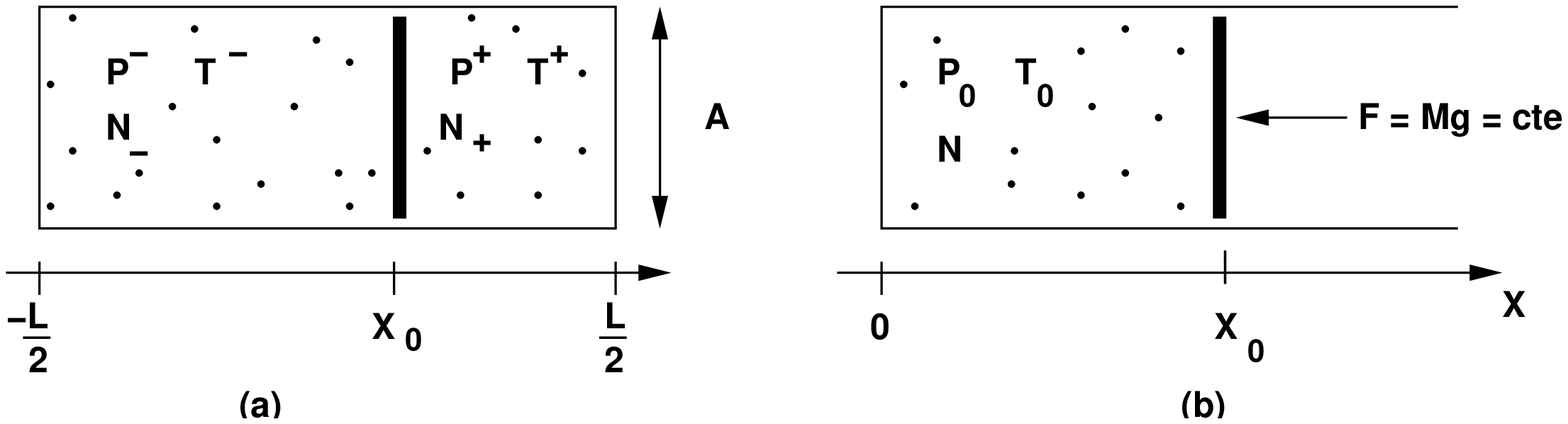}{\\ figure 1: (a) The   adiabatic piston,  (b)  The simple piston}
\end{center}

\vspace{0.1cm}

\noindent\noindent Experimentally the "adiabatic piston" has been used before $1940$ \cite{1}, to measure the ratio $\gamma = \frac{c_p}{c_v}$ of gases. From $1929$ until today this ratio has been mesured  \cite{2} by observing the damped "harmonic" motion of a simple piston under a constant external force $(fig. 1.b)$. The main result of the recent experiments is that one should distinguish between two regimes, corresponding to weak and strong damping with very different properties, $e.g.$

\begin{itemize}
\item for \underline{weak damping}  the frequency of oscillations is close to the  values computed assuming adiabatic oscillations.

\item for \underline{strong damping}, and small enough amplitudes,  the frequency correspond to isothermal  oscillations.
\end{itemize}

\noindent\noindent Since $1963$ the adiabatic piston became a controversial conceptual problem because it led to the conclusion that the two laws of thermo\underline{statics} are not sufficient to yield the final state $[3]$. In fact one is forced to solve the equations obtained from  thermo\underline{dynamics} to find the final state  $[4]$.

\noindent\noindent In  $1998$ we started to investigate this problem using a very simple microscopical model 
[$5-8$]. The system consists of $N$ point particles of mass $m$ in a cylinder of length $L$ and cross section $A$. The cylinder is divided into two compartments, containing respectively $N_-$ and $N_+$
particles $( N = N_- + N_+)$, by an "adiabatic piston" of mass $M \gg m$ orthogonal to the $x$-axis, $i.e.$ a wall with no internal degrees of freedom $(fig. 1.a)$. Initially the piston is fixed and the two gases are in thermal equilibrium. At time $t=0$ the piston is released and moves without friction, along the $x$-axis under the colisions with the gas particles. It is assumed that the gas particles move with constant velocity, except for purely elastic collisions on the boundaries of the cylinder and on the piston. The system is isolated and thus the energy $E$ is conserved.

Since the components of the particles velocities perpendicular to the piston play no role in the dynamic of the piston we assume that all the particles have velocities parallel to the 
$x$-axis and all distribution functions are independent of the transverse coordinates. Therefore if $v$ and $V$ denote the velocities of a particle and the piston before a collision, then under the collision
\begin{align}
v  \rightarrow v'= &2 V - v + \alpha  (v-V) \qquad\qquad V \rightarrow V'=  V +  \alpha  (v-V)\\
\text{where} \qquad \alpha =& \frac{2m}{M+m}
\end{align}
\vspace{-0.7cm}


\section{Numerical Results}\enlargethispage*{3\baselineskip}
\vspace{-0.2cm}
\indent\indent A systematic study of the time evolution of this model using numerical simulations have led to the conclusion that the evolution of the piston proceed essentially in two stages with totally different properties and time scales $[5-8]$.

\begin{enumerate}
\item In a \underline{first stage} caracterised by a time scale of order $\tau_1 = L\,\, \sqrt{\frac{M}{E_0}}
$
the evolution $X(t)$ of the piston has the following properties $(fig.2)$

\begin{enumerate}
\item It is \underline{deterministic} 
\item Let $\displaystyle{R_{\pm} = m \frac{N_{\pm}}{M}}$. For fixed $R_{\pm}$, the  evolution is \underline{independant of
 $ M$} for $M$ large enough ($e.g.  \,\,M> 10^4 m$). Moreover
\begin{itemize}
\item if  $R_{\pm} < 4$,  the   \underline{damping is weak}. The frequency of oscillations and the damping coefficient increase with $R_{\pm}$.
\item if  $R_{\pm} > 10$, the \underline{damping is strong}. The evolution is independant of $N_{\pm}$ and $M$.

\end{itemize}

\item The system evolves  \underline{adiabatically} ($i.e.$ without heat transfer through the piston) toward a state of "\underline{mechanical equilibrium}", where the pressure are equal $p^+=p^-=p_f$, but the temperature different.
In particular the temperature  of the gas increases under compression.

\end{enumerate}
\item In a \underline{second stage}, caracterised by a time scale of order
$
\tau_2 = \frac{M}m \,\,\tau_1
$,
which is very large if one considers realistic numbers, the evolution $X(t)$ is caracterised by $(fig.3)$
\begin{enumerate}\vspace{-0.1cm}
\item It is 
 \underline{stochastic}, with pressures $p^{\pm}$ fluctuating around $p_f$.
\item 
It depends strongly on $M$, but not on $N_{\pm}$. More precisely the following \underline{scaling} \underline{relation} is observed
$
X_M (t) = X(t/M)\nonumber
$
where $X(t)$ is independant of $M$ and $N_{\pm}$.
\vspace{-0.1cm}
\item The system evolves  \underline{with transfer of heat} through the piston toward a state of "\underline{thermal} \underline{equilibrium}" where $p^+ = p^-$  and $T^+= T^-$.

\noindent \noindent In particular in this stage the temperature of the gas decreases under compression.

\end{enumerate}

\end{enumerate}

\noindent \noindent Let us remark that some time after "mechanical" or "thermal"
equilibrium is reached, the piston starts oscillations which then disappear in time $(fig.2)$. These 
 oscillations reflect the fact that the velocity distributions of the gases are not maxwellian and they will be responsible for the approach to thermalisation with maxwellian distribution of velocities (see also $[10] $ were similar results have been obtained). 

\vspace{-0.2cm}
\section{Thermodynamic limit}
\enlargethispage*{1\baselineskip}\vspace{-0.1cm}
\indent\indent Although the second stage  is qualitatively understood and rather well described by Boltzmann Equation $ [8]$  (see $fig  3.b$), the same is not true for the first stage. To gain insight into truly adiabatic evolution, one can either study the simple piston $(fig. 1.b)$ where no heat is involved $[9]$, or consider the adiabatic piston in the thermodynamic limit $ [7]$ where $A \rightarrow \infty, N_{\pm} \rightarrow \infty, M \rightarrow \infty$ with fixed values for\vspace{-0.2cm} \begin{align} \,\,
L,\,\, R_{\pm} = m \frac{N_{\pm}}{M }, \,\, \text{and}  \,\,    \gamma = 2m \frac{A}{M+m}\end{align}
\noindent\noindent In this section we shall presents this second approach to the study of adiabatic evolution.
 Let us first consider a finite system, $i.e. A, N_{\pm}, M$
are all finite. We start from the Liouville Equation on the complete phase space for $N$ particles and the piston,with the dynamics defined by the elastic collisions (on the boundaries of the cylinder and the piston). Integrating  Liouville equations over all variables except one we obtain the equations for the one-particle distributions of the left and right particles $\rho^{\pm}(x,v;t)$ and of the piston $\Psi(X,V;t)$. It involves the two-particle distribution function  $\rho_{\pm, p }(x,v;X,V;t)$ for one particle and the piston:
\begin{align}
(\partial_t+v\partial_x)\rho^-(x,v;t) =&\delta (x+L/2) v \rho^-(x,v;t) +I^-\\
(\partial_t+v\partial_x)\rho^+(x,v;t) =&-\delta (x- L/2) v \rho^+(x,v;t) +I^+\\
\partial_t \phi (V;t)=& I
\end{align}

\vspace{-0.6cm}\begin{align}\text{where } \qquad\quad\qquad\qquad
\phi (V;t)= \int_{-\infty}^{+\infty}\, dX\, \Psi(X,V;t)\hspace{6.4cm}
\end{align}\vspace{-0.4cm}\hspace{8.2cm}
\begin{align}
I^-(x,v;t) &= \int_{v}^{+\infty}\, dV\, (V-v) \rho_{-, p }(x,v';x,V';t) 
-\int_{-\infty}^{v}\, dV\, (v-V)  \rho_{-, p }(x,v;x,V;t)\nonumber \\
I^+(x,v;t) &= \int_{-\infty}^{v}\, dV\, (v-V) \rho_{+, p }(x,v';x,V';t) 
-\int_{v}^{+\infty}\, dV\, (V-v)  \rho_{+, p }(x,v;x,V;t) \nonumber\\
I(V;t) &= A\,\, \bigg\{ \int_{-\infty}^{V}\,dv \,(V-v) \big[  \rho_{surf } ^- (v',V';t)   -
     \rho_{surf } ^+ (v,V;t)     \big]\nonumber \\
      &+\int_{V}^{+\infty}\, dv\, (v-V) \big[  \rho_{surf } ^+ (v',V';t)   -
     \rho_{surf }^+  (v,V;t)     \big]\nonumber
\bigg\}
\end{align}
 \begin{align}\text{ and } \qquad\qquad
 \rho_{surf } ^{\pm}(v;V;t) = \int_{-\infty}^{+\infty}\, dx   \rho_{\pm, p}(x,v;x,V;t) .\hspace{6.7cm}
 \end{align}
 \noindent In the thermodynamic limit $\alpha = \frac{2m}{M+m}$  is zero, and $Eq. 6$ can be written as 

 \begin{align}
\partial_t\phi (V;t)= - \gamma \frac{\partial}{\partial_V} \bigg\{ 
\int_{V}^{+\infty}\, dv\, (v-V)^2\rho_{surf } ^- (v,V;t)  
-  \int_{-\infty}^{V}\,dv \,(v-V)^2\rho_{surf } ^+ (v,V;t)  
\bigg\}.
  \end{align}
  \enlargethispage{1\baselineskip}
\noindent \noindent With the initial conditions we have considered one can  show that the solution is
 
\begin{align} 
  \Psi(X,V;t)&=\delta (X-X(t))\,\delta(V-V(t))
  \end{align}
 \begin{eqnarray}
\text{where}\quad\, \,\ \left\{
\begin{array}{l}

 \frac{d}{dt}X = V\\  \frac{d}{dt}V = 2m \frac{A}M \,\big[ F_2^- (V;t)  - F_2^+ (V;t) \big] 
\end{array}\right.
\end{eqnarray}
\begin{align} 
  F_2^- (V;t)=\int_{V}^{+\infty}\, dv\, (v-V)^2\rho_{surf } ^- (v;t) \qquad
   F_2^+ (V;t)=\int_{-\infty}^{V}\,dv \,(v-V)^2\rho_{surf } ^+ (v;t) \nonumber
  \end{align}
 \begin{align} 
 \rho_{\pm, p}(x,v;X,V;t) =&\rho^{\pm}(x,v;t)  \Psi(X,V;t)&\\ 
  \rho^{\pm}_{surf}(v;V;t) =&\rho^{\pm}(X(t),v;t) \delta(V-V(t))=  \rho^{\pm}_{surf}(v ;t)\delta(V-V(t))&
 \end{align}In conclusion in the thermodynamic limit $M\to \infty$, the evolution of the piston is deterministic given by the coupled equations $4-5$, $10-16$.  $Eq. 13$ can be written in the form 
 \begin{align}
  \frac{d}{dt} V = \frac{A}M\, (p^- - p^+)  - (\lambda^+(V)+\lambda^-(V))\,\,V
 \end{align} 
\noindent \noindent  where $p^{\pm}$ represent the \underline{pressure} exerted by the particles which are going to hit the piston and $\lambda^{\pm}(V)$ are strictly positive functions which represent the \underline{friction coefficients}:

 \begin{align}
  p^+ - p^- = 2m 
  \int_{0}^{\infty}\,dv \rho_{surf } ^+ (v;t) v^2
  -2m 
  \int_{-\infty}^{0}\,dv \rho_{surf } ^- (v;t) v^2
\end{align}  
 and
 \begin{align}
 \lambda^{\pm}(V)= \pm \frac2V m \frac{A}M \big[
 F_2^{\pm}(V) - F_2^{\pm}(0)\big]
 \end{align}  
 \noindent \noindent  Similary introducing the energy per unit area of the left/right compartment,
 \begin{align}
 e^{\pm}(t) = \int_{-\infty}^{+\infty}\,dv \rho^{\pm}(v;t) \frac12 m v^2, \qquad \rho^{\pm}(v;t) =\int dx 
  \rho^{\pm}(x,v;t)\nonumber
 \end{align}
 \noindent \noindent we obtain
 \begin{align} 
 \frac{d}{dt} e^{\pm}= \pm 2 m  F_2^{\pm}(V) V\nonumber
 \end{align}
 which shows that the change in the energy of the gases is entirely due to the work done by the piston: the evolution is therefore adiabatic in the thermodynamic limit.
 
 
 \section{Infinite cylinder $(L=\infty)$ and thermodynamic limit $(M=\infty)$ }
 \enlargethispage*{3\baselineskip}
 
\indent\indent If the cylinder is infinite and if the initial conditions are such that $|p^- - p^+ |$ is sufficiently small, we can expect that the recollisions of particles on the piston will be negligible. In that case the functions $ F_2^{\pm}(V)$ will be independant of $t$. Therefore the piston will evolve monotonically to the stationary state with constant velocity $\overline{V}$ solution of
 
 \begin{align} 
  \int_{\overline{V}}^{+\infty}\,dv \rho^{-}(v)\, (v-\overline{V})^2 =  \int_{-\infty}^{\overline{V}}\,dv \rho^{+}(v)\, (v-\overline{V})^2
 \end{align}
and the approach to the stationnary state is exponentially fast with time constant $\tau_0 = \frac1{\lambda}$, where $\lambda\cong  \lambda^+(V=0) + \lambda^-(V=0)$. Moreover the stationnary velocity does depend on the value $\frac{M}A$, and the time necessary to reach this stationnary state is proportionnal to $\big( m \frac{N}M\big)^{-1}$.

 
  \section{Thermodynamic limit for the piston $(M= \infty)$, but cylinder with finite length $(L<\infty)$}
 
\indent\indent In this section, to simplify the notation, we take $N_+= N_-= \overline{N}$ and $R= m \frac{\overline{N}}{M}$.
 In the case of a finite cylinder, recollisions of particles on the piston imply that $\rho_{surf}^{\pm} (v;t)$ depend explicitly on time and we have thus to study the complete  set of coupled equations.
 
 \noindent \noindent We remark first that if the friction coefficient $\lambda$ is sufficiently large the piston will reach almost immediately the constant velocity $ \overline{V}$ given by $(20)$ and this uniform motion should proceeed until the induced "shock wave" reaches the boundary. This expected result is well confirmed  by the numerical simulations for $R= m \frac{\overline{N}}{M}$ larger than $4$ ($fig. 2a$) .

 \noindent \noindent Since we can not solve the complete set of coupled equations we restrict ourselves to a qualitative discussion. If $p^+=p^- $ then $X(t) =X_0$ (recall that $(M= \infty)$). Therefore if the initial conditions are such that $ |p^+ - p^- | $ is small, we expect that $ |X(t) - X_0 | \ll 1 $ and we introduce the following \underline{\it assumption}
  \begin{align}  
  \rho_{surf}^-(t)&\cong\big(\frac{ N_-}A\big) \frac1{X(t)}= \rho^-(t);&\qquad  \rho_{surf}^+(t)\cong\big(\frac{ N_+}A\big) \frac1{L-X(t)}= \rho^+(t)\nonumber\\
   p^-(t)&\cong2 e^-(t)  \frac1{X(t)}=\rho^-(t) k_B T^-(t);&\qquad 
    p^+(t)\cong2 e^+(t)  \frac1{L-X(t)}=\rho^+(t) k_B T^+(t)\nonumber
   \end{align}
Moreover since $|V|$  is small we replace the friction coefficients $\lambda^{\pm} (V)$ by $\lambda^{\pm} (0)$ computed from maxwellian distributions, $i.e.$
 \begin{align} 
\lambda^{\pm} = \frac{A}M \sqrt{\frac{8}{\pi} k_B m} \rho^{\pm} \sqrt{T^{\pm} }
 \end{align}
With these approximations we have two constants of the motion
\begin{align} 
T^- + T^+ +\frac2{k_B R} V^2 = 2 T_0\\
 \sqrt{T^-}X - \sqrt{T^+} (L-X) = C.
\end{align}
We can thus find the equilibrium point, which is independant of ${\overline{N}}$ and $ M$:
\begin{eqnarray}
\left\{
\begin{array}{l}
\sqrt{X_f^3}- \sqrt{(L-X_f)^3}   =\sqrt{\frac{L}{2 T_0}} C \\ 
\displaystyle{T^-_f =  2 T_0 \frac{X_f}{L}}\\ 
\displaystyle{T^+_f =  2 T_0(1- \frac{X_f}{L})}
\end{array}\right.
\end{eqnarray}
Finally linearising the equations around the equilibrium point yield
\begin{align} 
\ddot{x}\,\,= &- w_0^2 x - \lambda \dot{x}\\
 w_0^2 =& \,6 R \frac{ k_BT_0}{X_f (L-X_f)}\\
  \lambda\,\, =& \sqrt{\frac{8}3\frac{R}{\pi}} \big[ 
  \sqrt{\frac{X_f}{L} } + \sqrt{1-\frac{X_f}{L} }\big ] w_0
\end{align}
The evolution is thus weakly damped if
\begin{align} 
R< R_{max} = \frac{3 \pi}{2}  \big[      \sqrt{\frac{X_f}{L} } + \sqrt{1-\frac{X_f}{L} }         \big ]\nonumber
\end{align}
\begin{align} \text{with period}\qquad\quad
\tau = \frac{2 \pi}{ w_0}   \frac1{\sqrt{1-\frac{R}{R_{max}}}}        
    \end{align}
    It is interesting to remark that $w_0$    coincide with the frequency for adiabatic oscillations $[2]$
  
 \vspace{-0.1cm}
  \section{Conclusions} \enlargethispage*{4\baselineskip}\vspace{-0.1cm}
\indent\indent Numerical simulations have shown that strictly adiabatic evolution can be obtained only in the thermodynamic limit for the piston. The study of a simple microscopical model has shown that in this limit  the evolution of the piston is deterministic,  caracterised by two regimes,weak and strong damping depending on whether $R= \frac{M_{gas}}{M}$  is small or large.
 For weak damping the frequency of oscillations coincide with the values obtained in numerical simulations. However the predicted value for damping is larger than the simulations by several orders of magnitude. For strong damping the initial uniform motion coincide with the simulation with the correct value for the friction coefficient.
 \\
 Using simplifying assumptions we obtained equations for the final equilibrium state in agreement with the numerical simulations; moreover one can show that the entropy of both gases increase in time, in agreement with the second law for adiabatic evolution.
 \\
 Finally it should be stressed that the adiabatic evolution that we have discussed is strictly hamiltonian,
 however in all cases we obtain a damping mechanism.
 This damping mechanism is still not well understood and the question of whether one can predict the final equilibrium state from fundamental principles remain an open problem.
  
 \vspace{-0.2cm} 
   \section*{Acknowledgment}\vspace{-0.1cm}
This paper is dedicated to H. Eugene Stanley in honor of his $60^{th}$  birthday. We are grateful to the "Fonds National Suisse de la Recherche Scientifique" for his financial support of this project.

 \twocolumn
 
\setcounter{figure}{1}

 \begin{figure}[H]

\centering
\subfigure[]{\label{e2}\includegraphics[width=6.1 cm,height=6.1 cm]{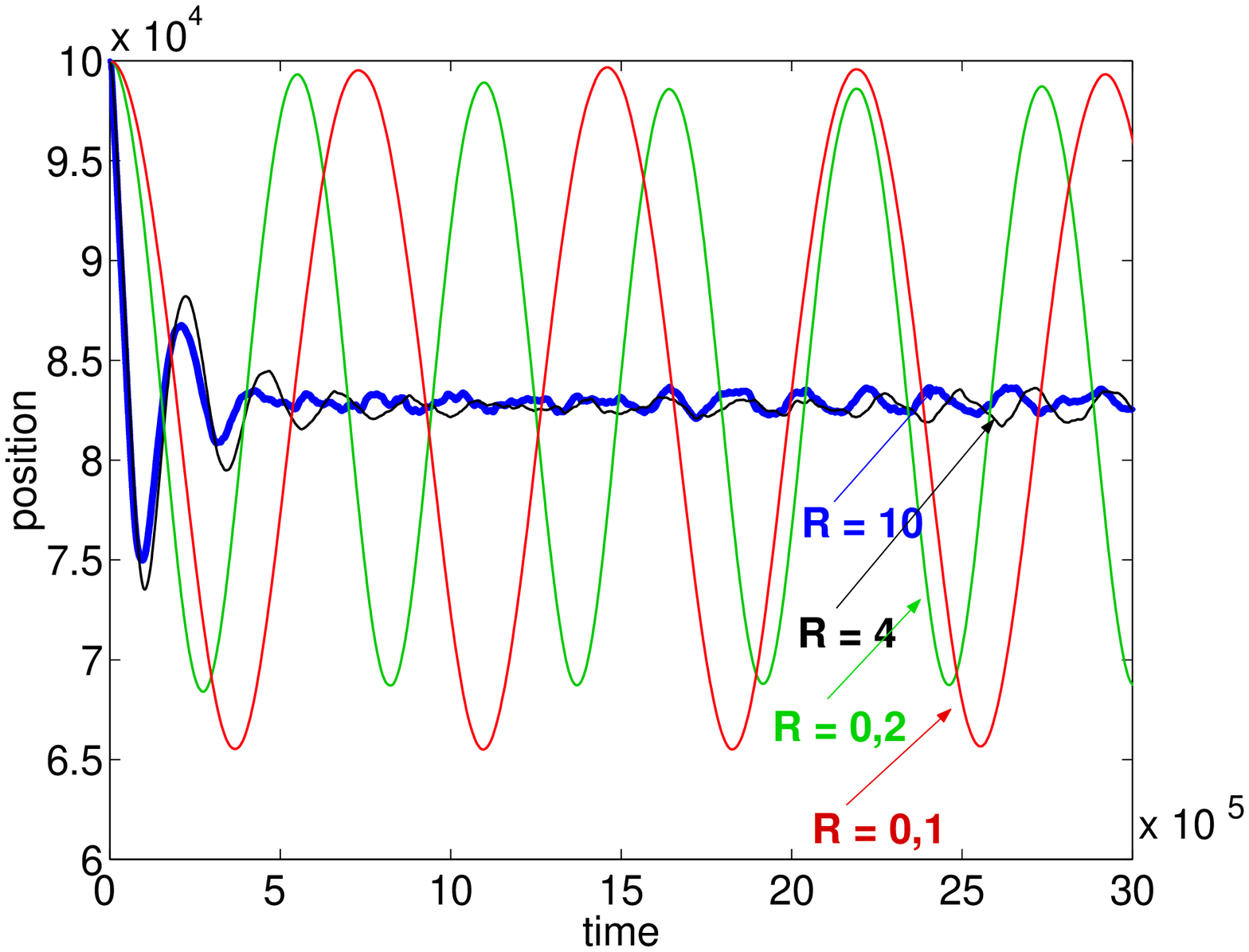}}\hspace{5mm}\vspace{-4mm}
\subfigure[]{\label{e1}\includegraphics[width=5.9 cm,height=5.7 cm]{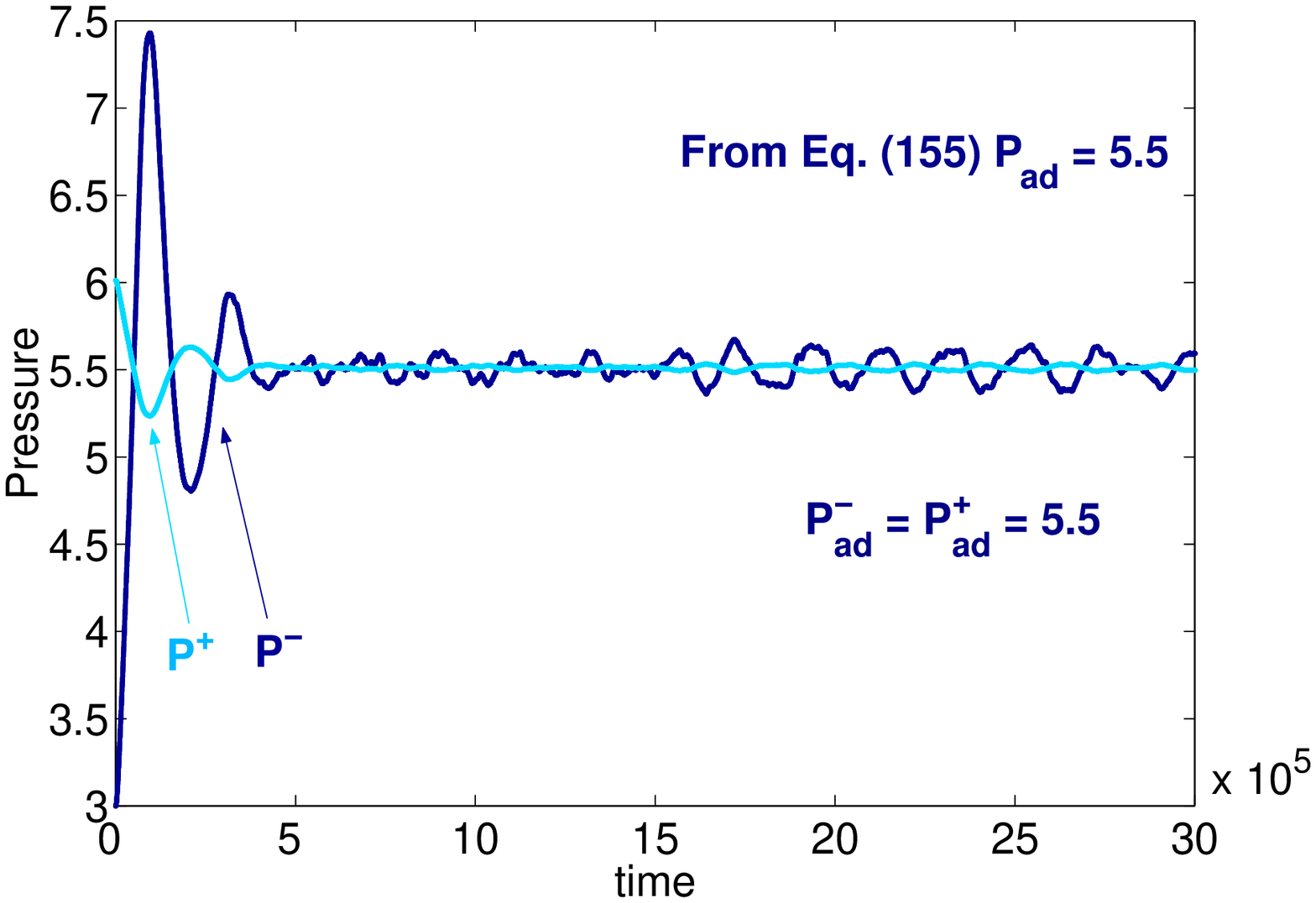}}
\vspace{3mm}
\hspace{1.8cm}\subfigure[]{\label{w}\includegraphics[width=5.9 cm,height=5.7 cm]{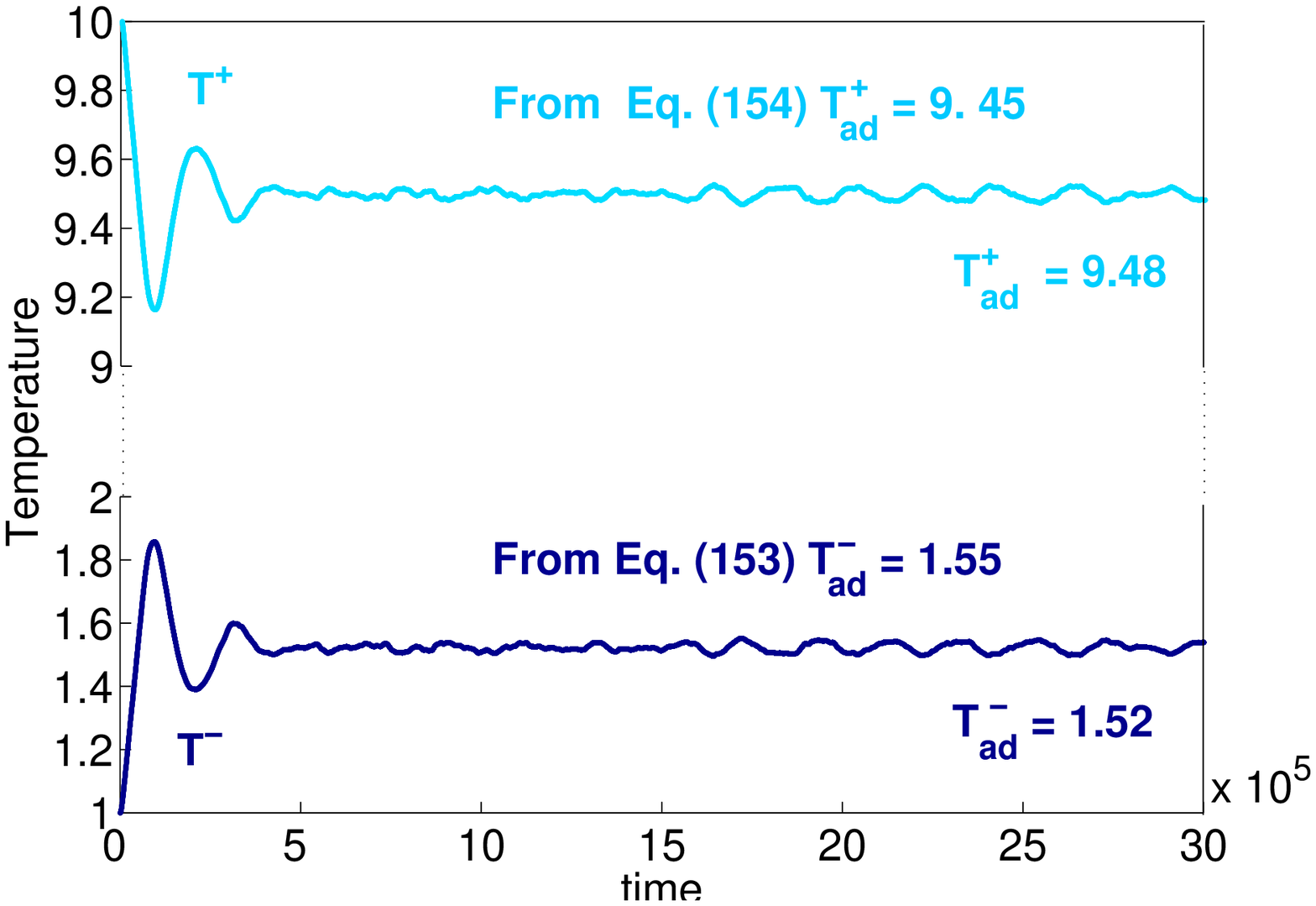}} \hspace{5mm}
\vspace{1mm}
\caption{ Adiabatic Evolution  }\label{E}\raggedright

$a)$ Position $X(t) $ for $ R = 0.1, 0.2, 4, M = 10^5$,\\ \hspace{3mm} $ R = 10 ,  M=30.000$
  \\$b)$ Pressure $p^{\pm} (t)$ for $R= 10, \,M=30.000$\\ $c)$ Temperature $T^{\pm} (t)$ for $R= 10, M=30.000$
\end{figure}

 \begin{figure}[H]

\centering

\subfigure[]{\label{ei}\includegraphics[width=6.5 cm,height=6.0 cm]{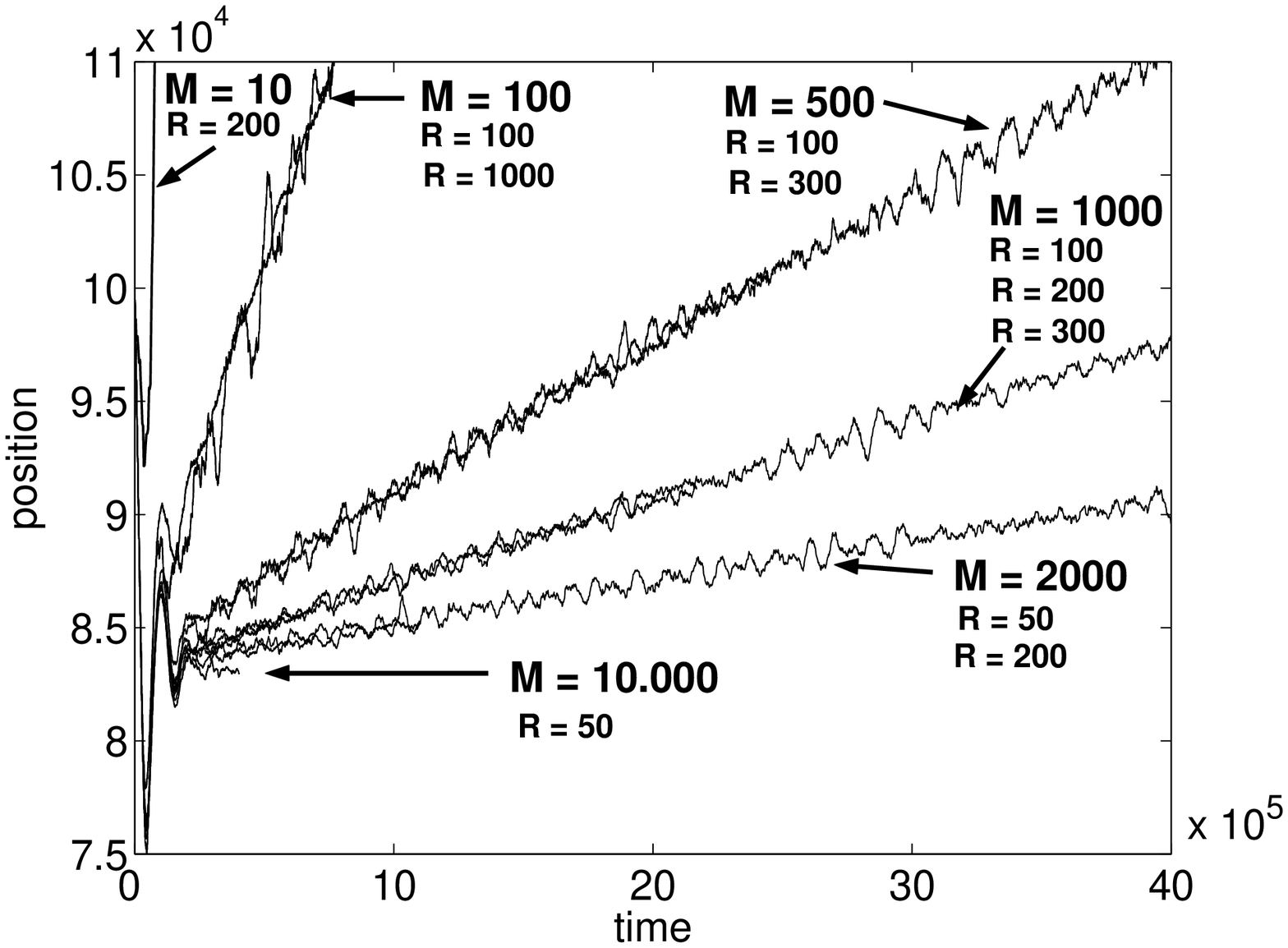}} \hspace{3mm}\vspace{-1mm}
 \hspace{6mm}\subfigure[]{\label{eoo}\includegraphics[width=6.1 cm,height=5.9 cm]{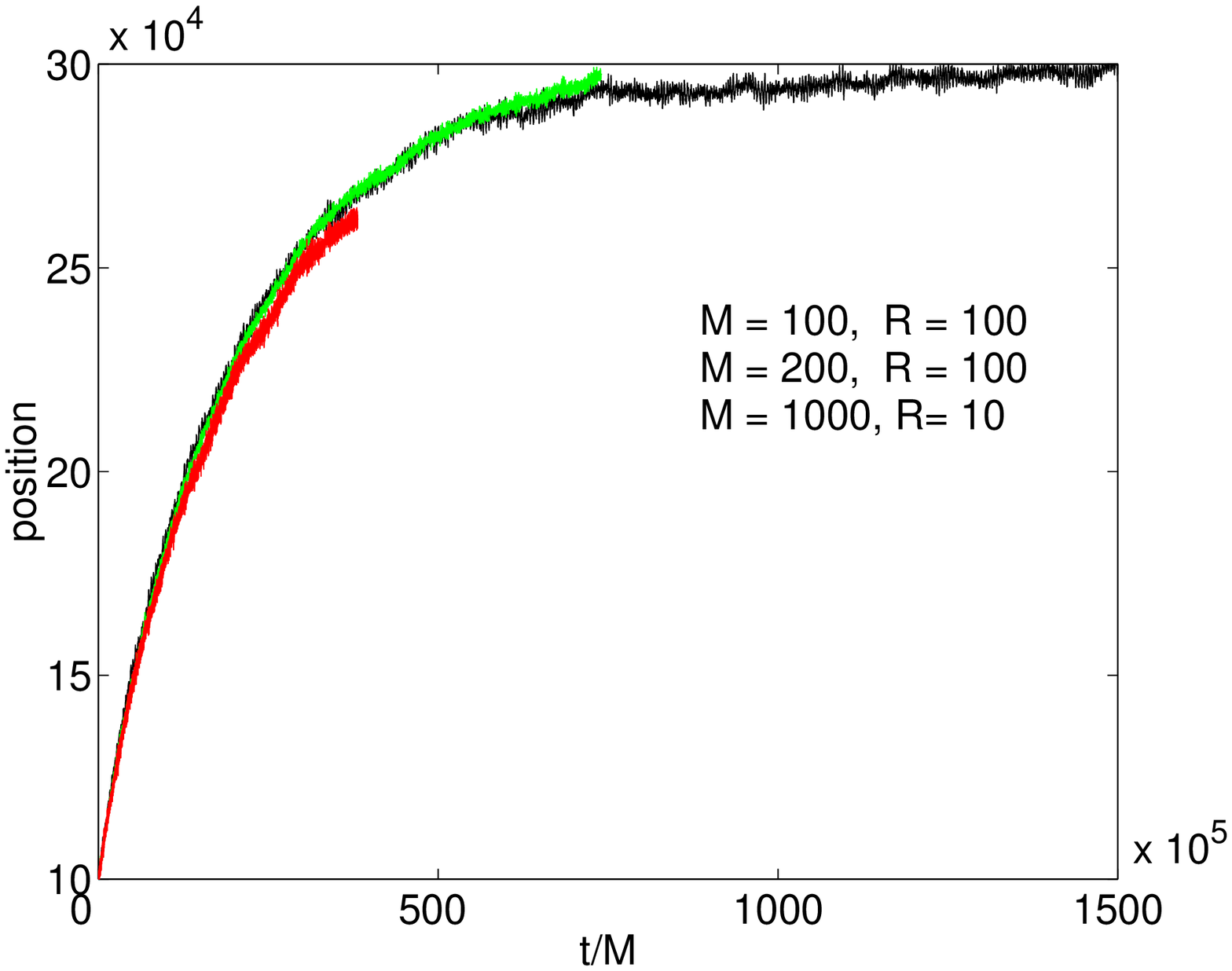}}
\vspace{-2mm} \hspace{2mm}
\subfigure[]{\label{e}\includegraphics[width=6.1 cm,height=5.8 cm]{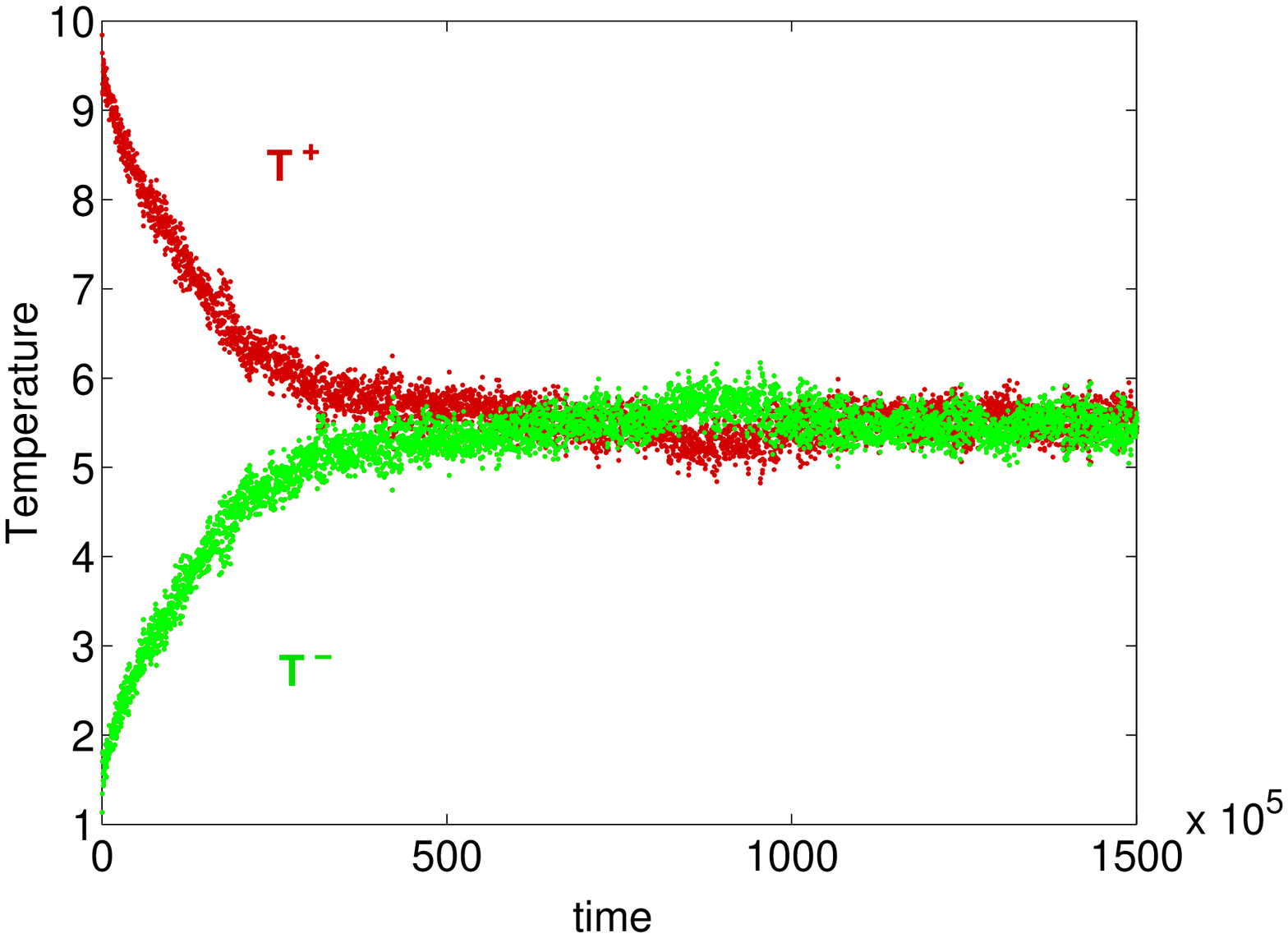}}\hspace{-1mm}

\vspace{-1mm} 
\caption{ Approach to thermal Equilibrium }\label{E}\raggedright
$a)$ Position \\ $b)$ Scaling property (Position)\\ $c)$ Temperature for $R= 10, M=100$
\vspace{0.4cm} 
\end{figure}

\vspace{-0.2cm} 
  
  
  
  
  
 
\begin{thebibliography}{99}

\bibitem{1} A. L. Clark and L. Katz, Can. J. Res. A {\bf 18}, (1940), 23-63
\bibitem{2} O. L. de Lange and J. Pierruis, Am. J. Phys. {\bf 68}, (2000) 265-270
\bibitem{3} H.B. Callen, {\it Thermodynamics}, John Wiley and Sons,
New York (1964)
\bibitem{4} Ch. Gruber, Eur. J. Phys. {\bf 20} (1999),  259-266.
\bibitem{5} Ch. Gruber, J. Piasecki, Physica A {\bf 268}  (1999), 258-265.
\bibitem{6} Ch. Gruber, L. Frachebourg, Physica A {\bf 272} (1999),
392-428.
\bibitem{7} Ch. Gruber, S. Pache, A. Lesne, ar Xiv: condomat / 0109542, 28 sept 2001, (submitted for publication).
\bibitem{8} Ch. Gruber, S. Pache, A. Lesne,  {\it Fluctuations and approach to thermal equilibrium for the enigmatic piston}, (in preparation).
 

\bibitem{9} G. P. Morris, Ch. Gruber, {\it Strong and weak damping  in the adiabatic motion of the simple piston},  preprint 2001. 

\bibitem{10} N. Chernov, J.L. Lebowitz, {\it Dynamics of a Massive piston in an ideal gaz: oscillatory motion and approach to equilibrium}, preprint 2001. 



  
  \end{thebibliography}
 \end{document}